# Specific Resistance of Pd/Ir Interfaces.

R. Acharyya, H.Y.T. Nguyen, R. Loloee, W.P. Pratt Jr., and J. Bass
Department of Physics and Astronomy, East Lansing, MI USA 48824

Shuai Wang and Ke Xia
State Key Laboratory for Surface Physics, Institute of Physics, Chinese Academy of Sciences,
Beijing, China

Abstract

From measurements of the current-perpendicular-to-plane (CPP) total specific resistance (AR = area times resistance) of sputtered Pd/Ir multilayers, we derive the interface specific resistance, $2AR_{Pd/Ir} = 1.02 \pm 0.06$ f$\Omega$m$^2$, for this metal pair with closely similar lattice parameters. Assuming a single fcc crystal structure with the average lattice parameter, no-free-parameter calculations, including only spd orbitals, give for perfect interfaces, $2AR_{Pd/Ir}(Perf) = 1.21 \pm 0.1$ f$\Omega$m$^2$, and for interfaces composed of two monolayers of a random 50%-50% alloy, $2AR_{Pd/Ir}(50/50) = 1.22 \pm 0.1$ f$\Omega$m$^2$. Within mutual uncertainties, these values fall just outside the range of the experimental value. Updating to add f-orbitals gives $2AR_{Pd/Ir}(Perf) = 1.10 \pm 0.1$ f$\Omega$m$^2$ and $2AR_{Pd/Ir}(50\text{-}50) = 1.13 \pm 0.1$ f$\Omega$m$^2$, values now compatible with the experimental one. We also update, with f-orbitals, calculations for other pairs.

In electronic transport with current-flow perpen-dicular to the layer planes (CPP geometry) of a metallic multilayer, the interface specific resistance AR (area A through which the CPP-current flows times the sample resistance R) is a fundamental quantity. In the past few years, measurements of AR have been published for a range of metal pairs [1-15]. Special interest focuses upon pairs M1 and M2 that have the same crystal structure and closely the same lattice parameters $a_o$—i.e., $\Delta a/a_o \leq 1\%$, since AR for such pairs can be calculated with no free parameters. That is, taking a given crystal structure and a common $a_o$ as known, the electronic structures for M1 and M2 can be calculated without adjustment using the local density approximation, and then $AR_{M1/M2}$ can be calculated without adjustment using a modified Landauer formula for either interfaces that are perfectly flat and not intermixed (perfect interfaces), or for interfaces composed of two or more monolayers (ML) of a 50%-50% random alloy (50-50 alloy) [16-21]. For all four such pairs (Ag/Au [18,20], Co/Cu [18,20,21], Fe/Cr [17,18], and Pd/Pt [12]) where experimental values of $2AR_{M1/M2}$ have been published, Table I shows that the previously calculated values for perfect and 2 ML thick alloyed interfaces of these pairs are not very different, and that the experimental values are generally consistent with both values to within mutual uncertainties.

In contrast, when $\Delta a/a_o$ is ~ 10%, the agreement between experiment and theory is only semi-quantitative—experiment and calculations differ by amounts as low as 50% to more than factors of two [11]. Moreover, a test [11] of decreasing the difference in lattice parameter from ~ 10% for Pd/Cu to ~ 5% for Pd/Ag and Pd/Au gave no improvement in agreement between experiment and theory. A subsequent comparison between calculations and experimental data on residual resistivities of a variety of impurities in different hosts showed that those calculations could be very sensitive to local strains [22]. Given these results, it seemed worthwhile to test a metal pair with a value of $\Delta a/a$ just beyond 1%, and simultaneously to test what should be improved calculations. In this paper, we examine the difference between experimental and calculated interface specific resistances for the pair Pd/Ir, for which we expected $(\Delta a/a_o) = (3.89 – 3.84)/3.89 = 1.3\%$ [23]. High angle x-ray studies of separately sputtered 200 nm thick Pd and Ir films gave $a_o(Pd) = 3.895 \pm 0.001$ Å, as expected, but $a_o(Ir) = 3.858 \pm 0.008$ Å, a bit larger than expected. Our measured values give $(\Delta a_o/a_o) = 1.0 \pm 0.2\%$. Low angle x-rays gave multilayer periodicities within 6% of the expected. We included such deviations in estimating experimental uncertainties.

Of the four pairs listed above, where good agreement between experiment and theory was found, three are continuously mutually soluble, and only one (Co/Cu) is barely mutually soluble at 295K [24]). Since Pd and Ir are only barely mutually soluble at 295K, our new study should also extend our understanding of such pairs.

The present study was double blind—no calculations were made prior to the measurements, and the experimenters at MSU and the theorists in China did not share results until both the measurements and calculations were independently completed. The results of both the measurements and the calculations are given in Table I. We will discuss them and their significance after describing our experimental and theoretical procedures.

Our samples are multilayers sputtered as described elsewhere [1]. To produce a uniform CPP current, the multilayers are sandwiched between crossed, 1.1 mm wide, Nb strips that become superconducting at our measuring temperature of 4.2K [1]. Experimentally, we determine the total specific resistance, AR, of a given multilayer by measuring its resistance R using a Superconducting-quantum-interference-device (SQUID) based bridge system [1], and measuring the area A ~ 1.2 mm$^2$ through which the CPP-current flows as described in [1]. To determine $AR_{Pd/Ir}$ experimentally, we use the technique of ref. [2]. This technique consists of measuring the total resistance AR as a function of the number of bilayers $n$ for a set of multilayers of the form Nb(100)/Cu(10)/Co(10)/[Pd(t)/Ir(t)]$_n$/Co(10)/Cu(10)/Nb(100) with equal thicknesses t of both Pd and Ir, and fixed total thickness $n(2t) = 360$ nm of the [Pd(t)/Ir(t)]$_n$. All thicknesses are in nm. In such samples, the total thickness of each metal, including Pd and Ir, is held fixed at a value independent of $n$. Thus, only the number of interfaces should change with $n$, until any finite thicknesses of the interfaces begin to overlap. So long as the interfaces don't overlap, the total AR should increase linearly with $n$ according to the equation:

$$AR = 2AR_{Nb/Co} + 2\rho_{Co}(10) + AR_{Co/Pd} + AR_{Co/Ir} + \rho_{Pd}(180) + \rho_{Ir}(180) + n(2AR_{Pd/Ir}). \quad (1)$$

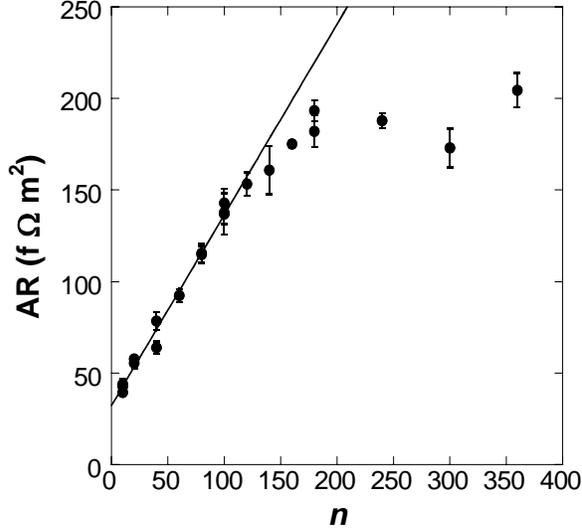

Fig. 1. AR vs $n$ for Pd/Ir multilayers. The line is the best linear fit up to $n = 120$.

This model subsumes into $2AR_{Pd/Ir}$ all contributions from the interfaces, including finite thickness effects. Our calculations will include a perfect zero-thickness interface and an alloyed interface, 2 ML thick.

If Eq. 1 applies, the slope of a linear plot of AR vs $n$ should give $2AR_{Pd/Ir}$, and the intercept AR($n=0$) should be given by the first six terms on the right-hand-side of Eq. 1. If t becomes so small that the layers completely overlap, the data should saturate to a value of the total AR representing 360 nm of a random 50%-50% Pd(Ir) alloy, plus those terms in the intercept part of Eq. 1 that do not involve Pd and Ir. Most of the six terms in the intercept are independently determined. $2AR_{Nb/Co} = 6 \pm 1$ f$\Omega$m$^2$ is the interface specific resistance between superconducting Nb and Co [25] (other studies [25] show that the Cu layers turn superconducting by the proximity effect and don't affect $2AR_{Nb/Co}$). We also measured the in-plane resistivities, $\rho_{Co} = 50 \pm 10$ n$\Omega$m, $\rho_{Pd} = 40 \pm 5$ n$\Omega$m, and $\rho_{Ir} = 110 \pm 15$ n$\Omega$m, of 200 nm thick Co, Pd, and Ir films sputtered at the same rates as for the multilayers. $AR_{Co/Pd} + AR_{Co/Ir}$ is the sum of the interface specific resistances for Co/Pd and Co/Ir. As we haven't measured these two terms, we assume, based on studies of other metal pairs [1-15], a total value $AR_{Co/Pd} + AR_{Co/Ir} = 1$ f$\Omega$m$^2$. This value is a small enough fraction of the total AR that uncertainties in this assumption are unlikely to be crucial. Adding these terms together gives a predicted intercept of $AR_{pred}(n = 0) = 35 \pm 5$ f$\Omega$m$^2$. We will compare this prediction with the extrapolation of our data.

Fig. 1 shows our data plotted as AR vs $n$. As expected, the data first grow linearly with $n$, and then begin to saturate, presumably as the finite thickness interfaces start to overlap. We are especially interested in deriving two quantities from Fig. 1.

(A) The slope of the linear part of the plot should give $2AR_{Pd/Ir}$. We show the best fit line up to $n = 120$, which is as far as a linear fit can go within the uncertainties of the data. From this line, and the range of fits consistent with the scatter of the data in Fig. 1, we derive a best estimate of $2AR_{Pd/Ir} = 1.02 \pm 0.06$ f$\Omega$m$^2$.

(B) The intercept of the linear part of the plot should give the sum of the first six terms on the right-hand-side of Eq. 1. Our best fit from Fig. 1 is $AR_{exp}(n=0) = 33 \pm 2$ f$\Omega$m$^2$. This value is consistent with our estimate above of $AR_{pred}(n=0) = 35 \pm 5$ f$\Omega$m$^2$.

In addition, if we assume that the approximate 'saturation' of AR for large $n$ is due to formation of a random 50%-50% PdIr alloy, we can estimate the resistivity of that alloy as $\rho(50\%\text{-}50\%\text{PdIr}) \approx 500$ n$\Omega$m. Unfortunately, neither of the usual sources of alloy resistivities [26,27] contains enough reliable data for PdIr to let us estimate this resistivity independently.

Having obtained an experimental value of $2AR_{Pd/Ir} = 1.02 \pm 0.06$ f$\Omega$m$^2$, we now turn to calculations to compare with this value. These assume completely diffuse scattering in the bulk metals, and no phase coherence in scattering from adjacent interfaces. As shown in ref. [16], these assumptions lead to Eq. 1 for either specular or diffuse interface scattering. As noted above, the calculations are based upon the local density approximation for the Fermi surfaces of Pd and Ir, and a Landauer formalism equation, as described in refs. [16-21]. We assume a single fcc lattice for the entire sample, with lattice parameter $a_o = 3.865$ Å, and calculate $2AR_{Pd/Ir}$ for (111) oriented planes. For a given assumed interfacial structure, the calculations can then be done without any adjustments. We do, however, test two analyses that differ in the maximum angular momentum $l(max)$, and thus the number of orbitals, included for the spin of each atom. The *spd* analysis, used for all prior calculations in Table I, assumes $l(max) = 2$ and involves 9 orbitals. The updated *spdf* analysis assumes $l(max) = 3$ and involves 16 orbitals. The first studies [18,20] used LMTO (linearized muffin tin orbitals). Later studies [19,21], and the present updated study, use MTO without linearization. Adding more orbitals, and not linearizing the MTO, should give more accurate potentials and thus more accurate band structures. For completeness, we also present in Table 1 new calculations applying the *spdf* analysis and MTO orbitals to prior metal pairs.

Although there is certainly interfacial intermixing, its precise extent is unknown. Thus, we calculated 2AR for both perfect interfaces and interfaces composed of 2 ML of a random 50%-50% alloy. Table 1 lists the results: $2AR_{Pd/Ir}(Perf)(a) = 1.21 \pm 0.1$ and $2AR_{Pd/Ir}(50-50)(a) = 1.22 \pm 0.1$. The listed uncertainties allow the calculated Fermi energies for Pd and Ir to deviate from experiment by $\pm 0.05$ eV [28]. The calculated *spd* values do not quite overlap with the experimental value. We checked that the calculations are not sensitive to the choice of average $a_o$ by recalculating for $a_o = 3.89$ Å or $3.84$ Å, the values for nominally pure Pd and Ir [23]. The changes were only ~ 2%. We also checked for sensitivity to inclusion of f-orbitals (*spdf*). Including them had a larger effect, reducing the calculated values to $2AR_{Pd/Ir}(Perf) = 1.10 \pm 0.1$ and $2AR_{Pd/Ir}(50-50) = 1.13 \pm 0.1$, now within mutual uncertainties of the experimental data. The similarities in 2AR for Perfect and 50-50 interfaces in Table 1 result from a competition between two opposite effects. Transport across a perfect interface requires conservation of $k_{||}$, the component of crystal momentum parallel to the interface. Scattering from the disordered defects in the 50-50 interface increases 2AR. But the disorder removes the $k_{||}$ constraint, reducing 2AR. For most pairs in Table I, these two effects roughly cancel.

In summary, we found an interface specific resistance for sputtered Pd/Ir multilayers of $2AR_{Pd/Ir}(exp) = 1.02 \pm 0.06$ f$\Omega$m$^2$. As listed in Table 1, we also calculated, with no adjustments, values of $2AR_{Pd/Ir}(calc)$ for a single fcc structure with a single lattice parameter, for the following conditions. (a) *spd* MTO calculations for both a perfect interface and an interface with 2 monolayers (ML) of a random 50%-50% Pd(Ir) alloy; and (b) *spdf* MTO calculations for perfect and 2 ML alloyed interfaces. For Pd/Ir, case (a) predictions are ~ 20% too large, which is not bad, considering there is no adjustability. Case (b) predictions agree with experiment to within mutual uncertainties. To conclude, *spdf* MTO calculations agree better with experiment for Pd/Ir than do *spd* MTO ones, change only a little from prior *spd* LMTO ones for Ag/Au, Co/Cu, and Fe/Cr, and are at about the limits of uncertainties of experiment for Pd/Pt.

Acknowledgments. This research was supported in part by the US NSF under grant DMR-08-04126. Ke Xia thanks NSF of China and MOST (No. 2006CB933000) of China.

References
1. S.F. Lee, et al., Phys. Rev. **B52**, 15426 (1995).
2. L.L. Henry, et al., Phys. Rev. **B54**, 12336 (1996).
3. N.J. List, et al., J. Magn. Magn. Mat. **148**, 342 (1995).
4. L. Piraux, et al., J. Magn. Magn. Mat. **156**, 317 (1996; *Ibid*, J. Magn. Magn. Mat. **159**, L287 (1996).
5. B. Doudin, et al., J. Appl. Phys. **79**, 6090 (1996).
6. J. Bass and W.P. Pratt Jr., J. Magn. Magn. Mat. **200**, 274 (1999)
7. W. Park, et al., Phys. Rev. **B62**, 1178 (2000).
8. A. Zambano, et al., J. Magn. Magn. Mat. **253**, 51 (2002).
9. K. Eid, et al., J. Appl. Phys. **91**, 8102 (2002).
10. H. Kurt, et al., Appl. Phys. Lett. **81**, 4787 (2002).
11. C. Galinon, et al., Appl. Phys. Lett. **86**, 182502 (2005).
12. S.K. Olsen, et al., Appl. Phys. Lett. **87**, 252508 (2005).
13. N. Theodoropoulou, et al., J. Appl. Phys. **99**, 08G502 (2006).
14. N. Theodoropoulou, et al., IEEE Trans. On Magn. **43**, 2860 (2007).
15. A. Sharma, et al., J. Appl. Phys. **102**, 113916 (2007).
16. K.M. Schep et al., Phys. Rev. **B56**, 10805 (1997).
17. M.D. Stiles and D.R. Penn, Phys. Rev. **B61**, 3200 (2000).
18. G.E.W. Bauer, et al., J. Phys. D **35**, 2410 (2002).
19. P.X. Xu et al., Phys. Rev. Lett. **96**, 17660 (2006).
20. K. Xia, et al, Phys. Rev. **B63**, 064407(2001).
21. K. Xia, et al, Phys. Rev. **B73**, 064420(2006).
22. P.X. Xu and K. Xia, Phys. Rev. **B74**, 184206 (2006).
23. N.W. Ashcroft and N.D. Mermin, "Solid State Physics", W.B. Saunders Publ., 1976.
24. M. Hansen, "Constitution of Binary Alloys, 2$^{nd}$ Ed.", McGraw-Hill, New York, 1958.
25. C. Fierz et al., J. Phys. Cond. Matt. **2**, 9701 (1990).
26. J. Bass, Metals, Electronic Transport Phenomena, Landolt-Bornstein, New Series **III**/15, 1982, Pg. 1.
27. K. Schroeder, "Handbook of Electrical Resistivities of Binary Metallic Alloys", CRC Press, 1983.
28. O.K. Anderson et al., Phys. Rev. **B 2**, 883 (1970)
29. S.F. Lee et al., J. Magn. Magn. Mat. **118**, L1, (1993).
30. T. Valet and A. Fert, Phys. Rev. **B48**, 7099 (1993).

| Metals (Struct.) | ($\Delta a/a$)% | 2AR(exp) (f$\Omega$m$^2$) | 2AR(Perf.) (f$\Omega$m$^2$) | 2AR(50-50) (f$\Omega$m$^2$) | 2AR(Perf.) (f$\Omega$m$^2$) | 2AR(50-50) (f$\Omega$m$^2$) |
|---|---|---|---|---|---|---|
| *Basis* | | | *spd* | *spd* | *spdf* | *spdf* |
| Ag/Au (fcc)(111) | 0.2 | 0.1 [2] | 0.09 [18,20] | 0.12 [18,20] | 0.09 | 0.13 |
| Co/Cu (fcc)(111) | 1.8 | 1.0 [6] | 0.9 [18,20,21] | 1.1 [18,20,21] | 0.9 [31] | 1.1 |
| Fe/Cr (bcc)(110) | 0.4 | 1.6 [8] | 1.9[18];1.5 [17] | 1.6 [18] | 1.7 | 1.5 |
| Pd/Pt (fcc)(111) | 0.8 | 0.28 ± 0.06 [12] | 0.30 ± 0.04 [12] | 0.33 ± 0.05 [12] | $0.40^{+0.03}_{-0.08}$ | $0.42^{+0.02}_{-0.04}$ |
| Pd/Ir (fcc)(111) | 1.3 | 1.02 ± 0.06 | 1.21 ± 0.1 | 1.22 ± 0.1 | 1.10 ± 0.1 | 1.13 ± 0.1 |

Table 1. Comparison of measured interface specific resistances, 2AR(exp), with calculations for perfect interfaces, 2AR(perf.), and for interfaces with 2 monolayers of a 50%-50% random alloy, 2AR(50-50), for metal pairs with the same crystal structures and closely the same lattice parameters. Both the experimental and calculated values of 2AR for the ferromagnetic/non-magnetic pairs, Co/Cu and Fe/Cr, are 2AR* as defined in refs. [29,30].